\documentclass[journal,twoside,web]{ieeecolor}
\usepackage{jsen}
\usepackage{cite}
\usepackage{amsmath,amssymb,amsfonts}
\usepackage{algorithmic}
\usepackage{graphicx}
\usepackage{textcomp}
\usepackage{wrapfig}
\usepackage{float}  % for placing figures to the right places
\usepackage{subcaption} %For sub-figures
\usepackage{hyperref}

\def\BibTeX{{\rm B\kern-.05em{\sc i\kern-.025em b}\kern-.08em
    T\kern-.1667em\lower.7ex\hbox{E}\kern-.125emX}}
\begin{document}
\title{Milli-RIO: Ego-Motion Estimation with Low-Cost Millimetre-Wave Radar}
\author{Yasin Almalioglu, Mehmet Turan, Chris Xiaoxuan Lu, Niki Trigoni, and Andrew Markham
\thanks{This work is funded in part by the NIST under Grant 70NANB17H185.}
\thanks{Yasin Almalioglu, Chris Xiaoxuan Lu, Niki Trigoni, and Andrew Markham are with the Computer Science Department, The University of Oxford, UK (e-mail: \{yasin.almalioglu, xiaoxuan.lu, niki.trigoni, andrew.markham\}@cs.ox.ac.uk). }
\thanks{Mehmet Turan is with the Institute of Biomedical Engineering, Bogazici University, Turkey (e-mail: mehmet.turan@boun.edu.tr).}}

\IEEEtitleabstractindextext{
%\begin{wrapfigure}[12]{r}{3in}%
%\includegraphics[width=2.5in]{figures/radar.jpeg}%
%\end{wrapfigure}%
\begin{abstract}
Robust indoor ego-motion estimation has attracted significant interest in the last decades due to the fast-growing demand for location-based services in indoor environments. 
Among various solutions, frequency-modulated continuous-wave (FMCW) radar sensors in millimeter-wave (MMWave) spectrum are
gaining more prominence due to their intrinsic advantages such as penetration capability and high accuracy.
Single-chip low-cost MMWave radar as an emerging technology provides an alternative and complementary solution for robust ego-motion estimation, making it feasible in resource-constrained platforms thanks to low-power consumption and easy system integration. 
In this paper, we introduce Milli-RIO, an MMWave radar-based solution making use of a single-chip low-cost radar and inertial measurement unit sensor to estimate six-degrees-of-freedom ego-motion of a moving radar. Detailed quantitative and qualitative evaluations prove that the proposed method achieves precisions on the order of few centimeters for indoor localization tasks.
\end{abstract}

}

\maketitle

\section{Introduction}
\label{sec:introduction}
%with the advent of integrated small and power-efficient mobile sensors.
\IEEEPARstart{R}{obust} ego-motion estimation for indoor environments has a variety of real-world applications ranging from emergency
evacuation to indoor robotics, and remains a challenging task.
In the last decades, with the advent of integrated small and power-efficient mobile sensors, various technologies have been adapted to this domain to investigate robust solutions.
A significant limitation of classical sensors such as vision or laser is that they are ineffective in visually degraded environments, e.g. glare, smoke and darkness \cite{fritsche2017fusion}.

\begin{figure}
\centering
\includegraphics[width=0.7\columnwidth]{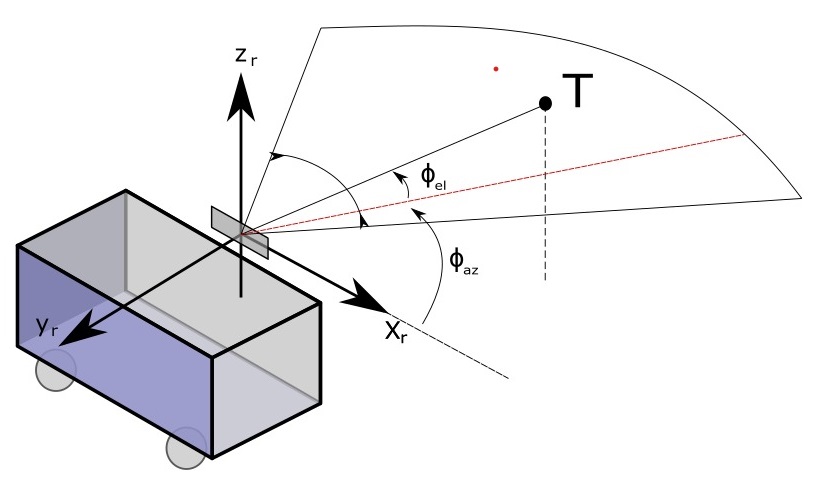}
\caption{\textbf{Millimetre-Wave (MMWave) radar system attached on top of a moving agent.} MMWave radar is a special class of radar technology that uses millimetre wavelength radio frequency (RF) signals. 
Our MMWave system operates at 76–81 GHz spectrum, resulting in an ability to detect movements on the order of few centimeters.} 
\label{fig:radar_overview}
\end{figure}

\begin{figure}
\centering
\includegraphics[width=\columnwidth]{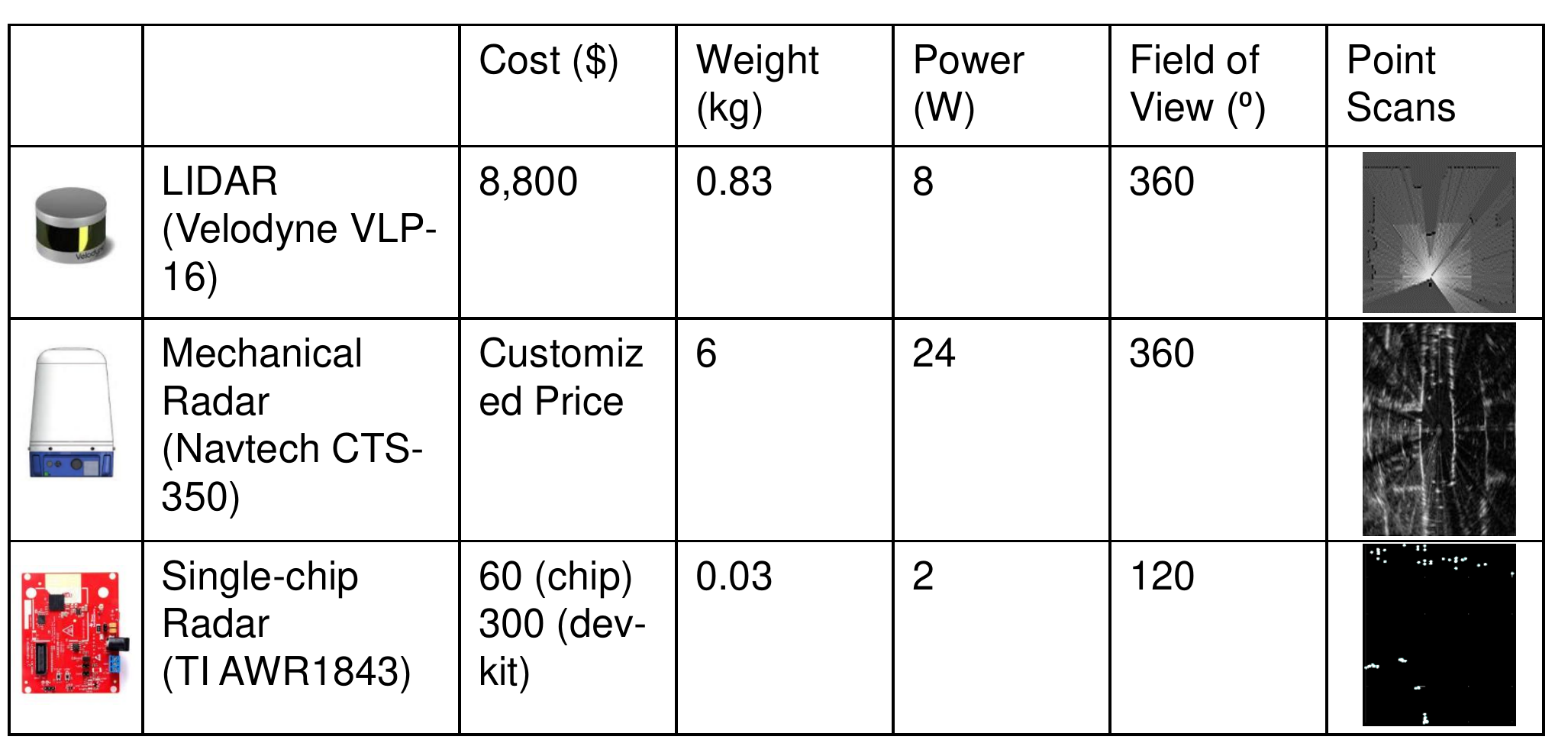}
\caption{\textbf{Sensor comparison.} Comparison of LIDAR, mechanical radar and single-chip radar, showing the features of a major commercial model for each sensor. Notably, compared with a LIDAR and a mechanical radar used in \cite{cen2018precise}, single-chip radar is much cheaper and lighter, but only provides few points within the field-of-view. White points in the scans correspond to detected objects.} 
\label{fig:sensor_comparison}
\end{figure}

%On the other hand, 
Radars provide robust and reliable perceptual information of the environment, which are immune to visual degradation.
Although radar systems were only used in the military area due to their bulky sizes and high costs in the early years, 
radar systems have been miniaturized and integrated onto printed circuit boards over recent decades thanks to the advance of high frequency integrated circuits.
With the recent advances in the integrated circuit and packaging
technologies, it is even possible to integrate a frequency-modulated continuous-wave (FMCW) radar system operating at a higher frequency millimeter-wave (MMWave) band (77 GHz) into a single chip with antenna-on-chip technologies.
Despite inherent issues such as high path loss, higher operation frequency and smaller wavelength do not only improve the sensitivity and resolution of radar systems but also make
radar systems further compact, extending the range of applications from military to commercial areas.
Low-cost MMWave radars thus provide a viable alternative (or complementary) solution for robust indoor ego-motion estimation to overcome the shortcomings of optical sensors (see Fig. \ref{fig:radar_overview} for an overview).
Figure \ref{fig:sensor_comparison} shows examples of the commercially available solutions used for ego-motion estimation, comparing low-cost radar in terms of cost, weight, energy consumption, field-of-view (FoV) and the detection density.

MMWave radar is extensively used in several domains such as the military (air and maritime surveillance, missile guidance, etc.), civil aviation (approach radar, surface movement radar) or remote sensing (planetary observation) \cite{mcmillan2006terahertz}. 
%Recent advances in radar technology and material science enabled a progressive adaptation of MMWave radars to smaller platforms in terms of dimension ($\sim20x30$cm), weight ($\sim200$gr), energy consumption ($\sim2$W)  and cost issues ($\sim$\$100), (see Fig. \ref{fig:radar_overview}). Their small size, low cost and fine accuracy make them suitable especially for portable low power applications \cite{charvat2014small}.
In recent years, several research groups have proposed MMWave radars as a solution for various mobile robot tasks such as navigation, localization and mapping.
In obstacle detection, MMWave radar is widely studied in automotive applications to detect moving and static targets (cars, pedestrians) \cite{hasch2012millimeter, zhao2019mid}. Several studies are proposed to investigate imaging capabilities of the radars for environment representation  \cite{brooker2015using} and 2D/3D simultaneous localization and mapping (SLAM) \cite{jaud2014boat}. MMWave radars are also fused with visual sensors for obstacle detection and map reconstruction, combining the robust depth detection ability of the radar in severe environmental conditions with a high spatial resolution of the visual sensors \cite{wang2011integrating}. 
However, these solutions involve bulky radar systems that provide dense measurements at the cost of increased physical size, power consumption and price of the system.
Thus, ego-motion estimation methods designed explicitly for single-chip, low-cost FMCW MMWave radars are needed to fully utilize the features of portable radar for indoor location-based services.

In this paper, we propose Milli-RIO, an ego-motion estimation method based on single-chip low-cost MMWave radar, which is complemented by an inertial measurement unit (IMU) sensor.  The main contributions of our method are as
follows:
\begin{itemize}
\item To the best of our knowledge, this is the first indoor ego-motion estimation approach using a single-chip low-cost MMWave radar sensor, making it effective for indoor applications in terms of size, cost and energy consumption.
\item We propose a new point association technique to match the sparse measurements of low-cost MMWave radar.
\item We propose a model-free motion dynamics estimation technique for unscented Kalman filter (UKF) using  Recurrent Neural Network (RNN).
\end{itemize}

As outline of the paper, Section \ref{sec:rel_work} presents the related work. Section \ref{sec:system} introduces the proposed ego-motion estimation method based on low-cost MMWave radar. The experimental setup is described in Sec. \ref{sec:exps_results}. The qualitative and quantitative results are presented in Section \ref{sec:exps_results}. Section \ref{sec:conclusion} concludes the study and gives future directions.

\begin{figure*}
\centering
\includegraphics[width=\textwidth]{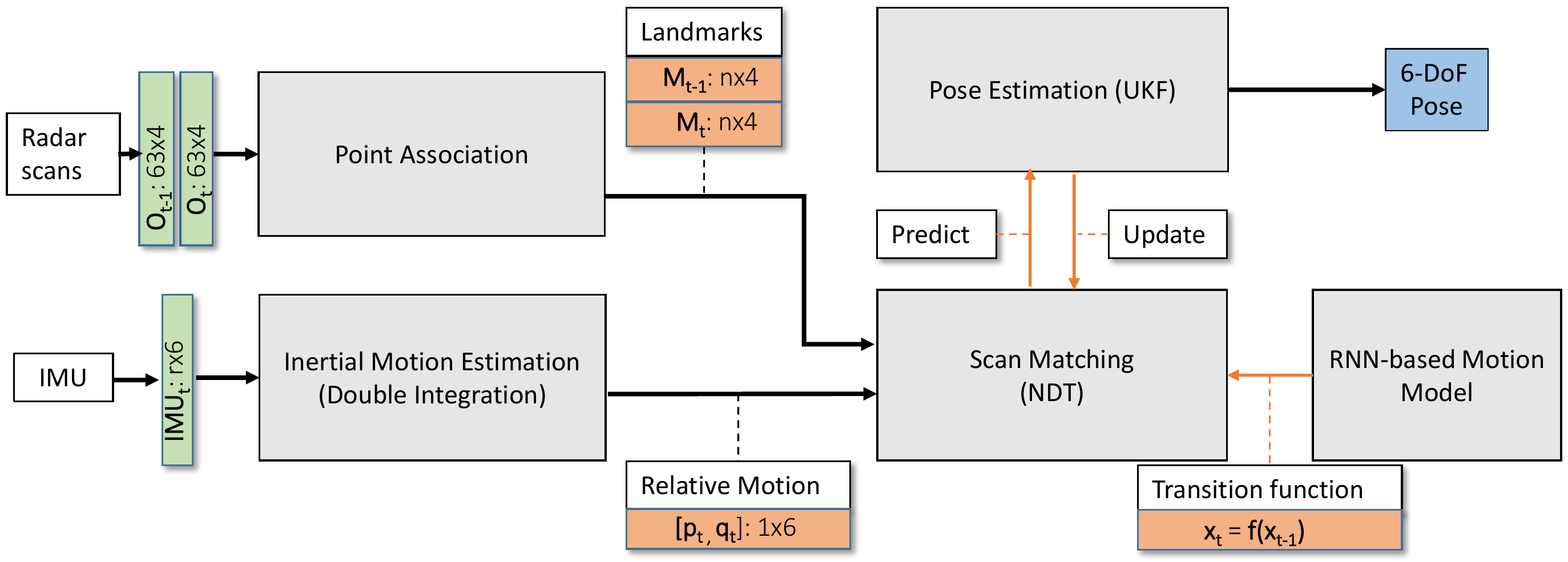}
\caption{\textbf{Ego-motion estimation workflow.} Raw MMWave radar point clouds are processed in the point association module to extract $n$ landmarks, where $n$ is the number of landmarks determined by the cost function. Scan matching module registers the landmarks using NDT scan matching algorithm, which uses a RNN-based transition model. In parallel, relative ego-motion is estimated from IMU readings using our inertial navigation system. Radar and IMU estimations are fused in the real-time pose estimation module using a UKF to regress the final 6 DoF pose values. 
Green, orange, gray and blue boxes represent inputs to the system, intermediate estimations, processing units, and outputs, respectively.} 
\label{fig:model_overview}
\end{figure*}

\section{RELATED WORK}
\label{sec:rel_work}
Radar typically has a wider, taller beam than the light detection and ranging (LIDAR) sensor, which makes scanning large volumes easier but results in lower bearing resolution and cluttered measurements. 
The longer wavelength of radars causes the radar echo to be reflected off multiple surfaces (such as the ground or walls) on its return trip to the antenna, known as the multipath effect. This effect delays the return of the signal and creates false targets further away than the real one, which is even more challenging in indoor environments due to walls, ceiling and floor reflections \cite{adams2012robotic}.
Thus, radar ego-motion estimation systems must be
robust to clutters and false-positives, and it
must demonstrate high precision despite low-resolution data.

Feature extraction is a fundamental task in radar ego-motion estimation systems. The traditional visual localization techniques such as amplitude grid maps are investigated in literature \cite{schuster2016landmark, werber2016interesting}. \cite{schuster2016landmark} uses the amplitude grid maps to transform the radar scans into grayscale images and applies SIFT and FAST feature extractions. \cite{werber2016interesting} studies the grid maps to find continuous areas using DBSCAN, MSER, and the connected components. 
The radar-specific solutions utilize data distortion, which is used as sources of information to estimate vehicle displacement \cite{vivet2013localization}. Another technique exploits spatiotemporal continuity in radar scans inferring the appearance of landmarks by determining the radar noise characteristics \cite{jose2005augmented}. In 2D radar scan processing, the accurate range information calculated with the highest power return per azimuth eliminates the need for a filter, which can potentially discard relevant information \cite{marck2013indoor}. 
To combine visual and radar sensors, \cite{schuster2016landmark} pairs radar and visual landmarks with similar feature descriptors. Vision-radar fusion approaches use radar occupancy grids to associate both sensor measurements \cite{schuster2016robust}. Feature descriptors work well for images that contain complex and high-density information. However, they are unable to create useful feature descriptions from radar scans that characteristically have significant noise and sparse data. Multi-sensor fusion techniques provide an alternative to feature-based radar odometry. The fusion methods use odometry information from additional sensors to transform the incoming radar landmark point cloud and register it to an existing landmark map. They usually make use of nearest neighbor point matching \cite{schuster2016robust} and Monte Carlo methods to derive a solution from probabilistic weights \cite{deissler2010uwb}. The relative motion is estimated using the data association between the radar point cloud and map, which is then fused to the first odometry readings. Although existing multi-sensor fusion methods are promising, they make use of sensors that already provide highly accurate odometry results. 

In radar-based ego-motion estimation systems, the extracted features are processed in the data association step, which is frequently achieved by a scan matching algorithm that tracks shared features across consecutive radar scans. 
The iterative closest point (ICP) approach is typically used for scan matching to iteratively align the radar point clouds until the predefined termination criteria are met \cite{ward2016vehicle}, which is too sensitive to outliers. In \cite{chandran2006motion}, the researchers developed a quantitative function describing the quality of the map created by superimposing radar point clouds according to the unknown motion parameters. 
An innovative technique well suited for high velocities utilizes the radar scan distortions that are often a drawback of mobile radar systems to eliminate the high-velocity effects using an extended Kalman filter \cite{vivet2013localization}. Other scan matching algorithms operate directly on the radar outputs instead of extracting landmarks. The Fourier-Mellin transform enables efficient computation of the vehicle's rotation and translation from the entire radar output \cite{checchin2010radar}. The Doppler radar returns the position and speed of the objects around, and the vehicle motion is easily computed relative to the surrounding objects given a sufficient amount of radar scans \cite{kellner2014instantaneous}. Both methods are hampered by heavy preprocessing.
\cite{cen2018precise} proposes a data association method that requires power-range spectra and the extracted landmarks to match similar geometries within two radar scans. The technique requires dense radar scans that already contain visible shapes and patterns, and is not practical for sparse radar measurements.

The present works above for radar-based ego-motion estimation are developed for mechanically rotating radars, which already provide a dense and full field-of-view (FoV) of the environment. 
The mechanical radars are bulky, expensive and power-hungry sensors, which is incompatible with the requirements of portable indoor location-based services \cite{brena2017evolution} (see Table \ref{fig:sensor_comparison}). 
For indoor navigation and mapping, single-chip radars operating at MMWave frequencies have the following advantages over the other radars: a) reduction in size and mass of the radar allows developing small ultra-lightweight radars suitable for portable indoor systems, and b) millimeter wavelength allows detection of relatively smooth wall surfaces at oblique angles of incidence. 
However, challenges of single-chip radars such as sparse measurement and limited FoV require specific solutions.
In this paper, we present a novel and robust motion estimation approach for indoor localization tasks based single-chip MMWave radar, complemented by the IMU sensor to eliminate deficiencies of both sensors such as biases in IMU output, noises and sparse measurements in radar scans.

\section{Millimetre-Wave Radar Based Ego-Motion Estimation}
\label{sec:system}
Frequency modulated continuous wave (FMCW) MMWave radar has the ability to simultaneously measure the range and
relative radial speed of a target point. 
Milli-RIO is an ego-motion estimation system that exploits
the unique properties of single-chip MMWave radar. It transmits an RF signal and records reflection from a target point that is collected in a point cloud.
It then calculates ego-motion by registering the generated sparse point cloud, which uses IMU as an auxiliary sensor to improve registration performance.
In this section, we describe the principles of MMWave radar and present the proposed MMWave radar-based point association and ego-motion estimation algorithms. Moreover, we explain the details of the RNN-based motion model used in the joint MMWave radar-IMU ego-motion estimation. 

\subsection{Principles of MMWave Radar}
FMCW MMWave radar uses a linear ‘chirp’ or swept frequency
transmission that is characterized by a bandwidth $B$, start frequency 
$f_c$ and duration $T_c$. 
A mixer in the radar front-end computes the frequency
difference between the transmitter and the receiver.
the distance between the object and the
radar is calculated from an Intermediate Frequency (IF) as:
\begin{equation}
d= \frac{f_{IF^{C}}}{2S},
\end{equation}
where $c$ is the speed of light, $f_{IF}$ is the
frequency of the IF signal, and $S=B/T_c$ is the frequency slope of
the chirp.
Each peak of the FFT result on the IF signal represents an obstacle at a corresponding distance.

The angle of arrival (AoA) is estimated using the slight difference in phase of the received signals and emitted chirp signal. For a pair of antennas, AoA is calculated as:
\begin{equation}
\theta=\sin ^{-1}\left(\frac{\lambda \omega}{2 \pi d}\right)
\end{equation}
where $\omega$ is the phase difference nad $\gamma$ is the wavelength. 
The average result from the receiver pairs gives the final AoA, which decreases with $|\theta|$.

\begin{figure}
\centering
\includegraphics[width=\columnwidth]{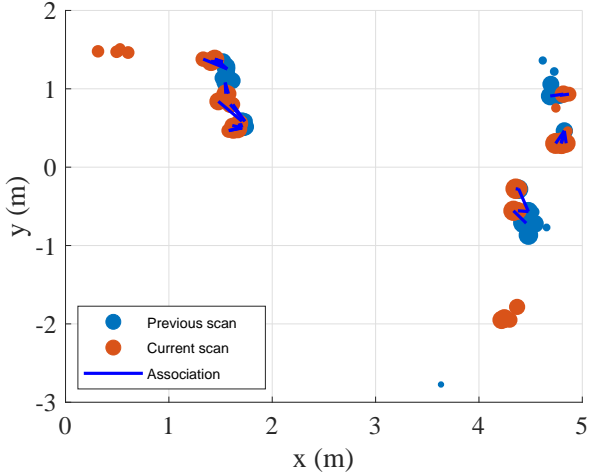}
\caption{\textbf{Example of point association for MMWave radar scans.} The marker sizes are proportional to the power intensity values. The proposed point association method finds the best matches, eliminating the low-intensity points and the ghost objects. It also penalizes the objects in the current scan that have no counter-part in the previous scan. The radar moves at approximately $0.50$ m/s. Positions are relative to radar position.}
\label{fig:data_assoc}
\end{figure}

\subsection{Point Association}
\label{subsec:data_assoc}
Milli-RIO achieves robust point correspondences across consecutive scans using high-level information in the radar output. The proposed point association technique seeks to find the largest subset of points that have the same landmark descriptors. 

Given two consecutive sets of detected radar points $\mathcal{O}^{t}=\{\mathbf{o}_{1}^{t},\mathbf{o}_{2}^{t},\ldots,\mathbf{o}_{N}^{t}\}$
at time $t$, and $\mathcal{O}^{t-1}= \{\mathbf{o}_{1}^{t-1},\mathbf{o}_{2}^{t-1},\ldots,\mathbf{o}_{M}^{t-1}\}$ at time $t-1$ in Cartesian coordinates, we aim to estimate the rigid-body transformation represented in homogeneous coordinates between the current and the previous scans.
This problem of rigid transformation estimation between two sets of points is well researched, especially in the robotics and computer vision field, where it is often solved with the iterative closest point (ICP) algorithm \cite{besl1992method}.
However, points detected by the radar are not stable over time due to the noise and stronger reflectors within the FoV. 
Thus, a point appearing in the previous scan may not be present at the current scan. In addition, the number of detected points is low (lower than 100 points per frame), which is typically around 1000 times less than a full FoV scan of a mechanical radar or LIDAR \cite{fritsche2016radar, barnes2019oxford}. 
ICP performs poorly on this type of data due to these challenges. 
We propose a new technique in this paper, in which point association and motion estimation are performed separately using radar specific features.
 
The proposed technique performs point association using not only individual unary landmark descriptors but also high-level descriptors between landmarks such as signal intensity and point cloud displacement.
This approach reduces the likelihood of an individual point having the same set of descriptors as another. 
Moreover, the signal intensity is not dependent on the exact position and orientation of the point cloud, making large disparities in pose inconsequential. 
In addition, the proposed point association algorithm is not constrained to have a good initial estimate of the relative pose, enabling feature extraction from point clouds captured at arbitrary times without any \textit{a priori} map representation.
These properties provide reliable matches for our sparse landmark sets.

%Data Association
We pose the problem of point association as a linear sum assignment problem. Given two point sets $\mathcal{O}^{t}$ and $\mathcal{O}^{t-1}$, the objective is to find a complete assignment while minimizing a cost:
\begin{equation*} 
\min\sum\limits_{i}\sum\limits_{j}D_{i, j}A_{i, j} \tag{2} 
\end{equation*}
where $D_{i, j}$ is the score of matching point $i$ in $\mathcal{O}^{t}$ with point $j$ in $\mathcal{O}^{t-1}$ and $A$ is a binary matrix such that $A_{i, j}=1$ if point $\mathbf{o}_{i}^{t}$ is matched with $\mathbf{o}_{j}^{t-1}$.
We find the optimal alignment using the  Munkres algorithm \cite{munkres1957algorithms}.

We construct the similarity score matrix $\mathbf{D}$ as follows:
\begin{equation*} 
D_{i, j}=\begin{cases} \frac{1}{1+ dst(\mathbf{o}_{\mathbf{i}}^{t},\mathbf{o}_{\mathbf{j}}^{t-1})} & \text{if}\ \varphi(\mathbf{o}_{\mathbf{i}}^{t},\mathbf{o}_{\mathbf{j}}^{t-1})==True\\ 0 & \text{otherwise} \end{cases} \tag{3} 
\end{equation*}
where the unary landmark descriptor $dst(\mathbf{o}_{\mathbf{i}}^{\mathbf{c}},\mathbf{o}_{\mathbf{j}}^{\mathbf{p}})$ is the squared Euclidean distance between the points  $\mathbf{o}_{\mathbf{i}}^{t}$ and $\mathbf{o}_{\mathbf{j}}^{t-1}$, measuring that correctly identified landmarks have the same spatial and dynamic features in any two radar scans.
$\mathbf{D}$ is diagonally dominant for the optimal set of matches, making the overall pair-wise similarity maximum.
$\varphi(\mathbf{o}_{\mathbf{i}}^{t},\mathbf{o}_{\mathbf{j}}^{t-1})$ is a policy function that defines the high-level landmark descriptors to eliminate improbable associations caused by ghost reflections and noise. $\varphi(\mathbf{o}_{\mathbf{i}}^{t},\mathbf{o}_{\mathbf{j}}^{t-1})$ returns $\mathit{False}$ if two points cannot be associated.
Given the assumptions that (a) the radar moves forward with a given maximum speed, (b) the lateral translation is low, and (c) the landmarks have minimum signal intensity value,
we define $\varphi(\mathbf{o}_{\mathbf{i}}^{t},\mathbf{o}_{\mathbf{j}}^{t-1})$ as follows:
\begin{equation} 
\varphi(\mathbf{o}_{\mathbf{i}}^{t},\mathbf{o}_{\mathbf{j}}^{t-1})=\begin{cases} False & \text{if}\ dst(\mathbf{o}_{\mathbf{i}}^{t},\mathbf{o}_{\mathbf{j}}^{t-1}) > MaxValue\\ 
False & \text{if}\ o_{j}^{t}\vert_{x}-o_{i}^{t-1}\vert_{x} < 0\\ 
False & \text{if}\ (o_{j}^{t}\vert_{y}-o_{i}^{t-1}\vert_{y})^{2} > MaxLateral\\ 
False & \text{if}\ I(o_{j}^{t}) < MinIntensity\\
True & \text{otherwise} \end{cases}
\end{equation}
where $o_{j}^{p}\vert_{x}$ and $o_{j}^{p}\vert_{y}$ respectively denote the longitudinal component (along the $x$ axis) and the lateral component (along the $y$ axis) of $\mathbf{o}_{\mathbf{j}}^{\mathbf{p}}$. $\mathit{MaxValue}$ denotes the maximum distance, and $\mathit{MaxLateral}$  the maximum lateral displacement of the ego vehicle between two iterations. 
$I(o_{j}^{t})$ denotes the signal intensity of point $o_{j}^{t}$, which is conditioned on the required minimum signal intensity $\mathit{MinIntensity}$. 
Note that $\varphi(\mathbf{o}_{\mathbf{i}}^{t},\mathbf{o}_{\mathbf{j}}^{t-1})$ could also be defined with additional information, for example using the previous results of the motion estimation or using information from the odometers.

The greedy method iteratively collects satisfactory matches into the set $\mathcal{M}$. On each iteration, it evaluates the remaining valid matches and calculates the score values.
The remaining point pairs $(i,j)$ with a score $D_{i,j}>d_{\tau}$ are collected in a set of matches $\mathcal{M}$ for a given score threshold $d_{\tau}$.
An example of point association given in Fig. \ref{fig:data_assoc} shows that the point association is coherent despite the noisy detections, where the ego speed is around $0.50$ m/s.
%A pseudo-code for data association is shown in Fig. XX.

\subsection{Relative Motion Estimation}
%The proposed motion estimation method performs the data association on $\mathcal{M}$ (Section \ref{subsec:data_assoc}), which is invariant to arbitrary rigid body transformations in terms of distance and rotation.
In the relative motion estimation module of the proposed system, we estimate the sensor
trajectory by iteratively applying the normal distributions
transform (NDT) scan matching technique \cite{magnusson2007scan} to find the rigid body motion given the set of corresponding points $\mathcal{M}$. 
NDT is shown to have a better performance than other
scan matching algorithms, such as ICP, in terms of both reliability and processing speed \cite{magnusson2009evaluation}.
We can estimate the sensor ego-motion by iteratively
applying a scan matching algorithm.
However, the performance of any scan matching algorithm is affected by the number of point correspondences between two sets, which might fail due to large displacements caused by rapid motions. 
To
deal with this problem, we integrate angular velocity data provided by the IMU sensor to the NDT scan matching algorithm using UKF \cite{wan2000unscented}.
The pipeline of our method is demonstrated in Fig. \ref{fig:model_overview}.

We define the sensor state to be:
\begin{equation}
\mathbf{x}_t = [\mathbf{p}_t, \mathbf{q}_t, \mathbf{v}_t, \mathbf{b}_t^a]^T,
\end{equation}
where, $\mathbf{p}_t$ is the position, $\mathbf{q}_t$ is the rotation quaternion, $\mathbf{v}_t$ is the velocity, $\mathbf{b}_t^a$
is the bias of the angular velocity of the
sensor at time $t$. Assuming a transition function $f(\cdot)$ for
the sensor motion model and constant bias for the angular
velocity sensor, the system equation for predicting the state
is defined as:
\begin{equation}
\label{eq:predict_x}
\mathbf{x}_t = [\mathbf{p}_{t-1}+f(\mathbf{x}_{t-1}), \mathbf{q}_{t-1}.\Delta \mathbf{q}_{t}, \mathbf{v}_{t-1}, \mathbf{b}_{t-1}^a]^T,
\end{equation}
where $\Delta \mathbf{q}_{t}$ is the rotation during $t-1$ and $t$. 
The rotation is given by:
\begin{equation}
\Delta \mathbf{q}_{t} = \left[ 1, \frac{\Delta t}{2}\mathbf{a}_{t}^{x'}, \frac{\Delta t}{2}\mathbf{a}_{t}^{y'}, \frac{\Delta t}{2}\mathbf{a}_{t}^{z'} \right]^T,
\end{equation}
where $\mathbf{a}_{t}^{'}=\mathbf{a}_{t}-\mathbf{b}_{t-1}^a$ is the bias-compensated angular velocity.

The motion estimation module uses Eq. \ref{eq:predict_x} and UKF to predict the sensor pose, the estimated $\mathbf{x}_t$ and $\mathbf{q}_t$ being the initial guess of the sensor pose. 
We iteratively apply NDT on the set $\mathcal{M}$ to register the observed point cloud into the global map.
Then, the system corrects the sensor state using
sensor pose estimated by the scan matching $\mathbf{z}_t = [\mathbf{p}_t^{'}, \mathbf{q}_t^{'}]^T$.
The observation equation of UKF is defined as:
\begin{equation}
\mathbf{z}_t = [\mathbf{p}_t, \mathbf{q}_t]^T.
\end{equation}

We normalize $\mathbf{q}_t$ in the state vector after
each prediction and correction step of UKF to avoid norm changes due to unscented transform and
accumulated calculation error. It is worth mentioning that we
also implemented pose prediction, which takes acceleration
into account, as well, but the estimation performance deteriorates due to strong acceleration noise and constant bias. Thus, we omit the acceleration update step from Milli-RIO.

\begin{figure}
\centering
\begin{subfigure}[t]{0.45\columnwidth}
\includegraphics[width=\textwidth]{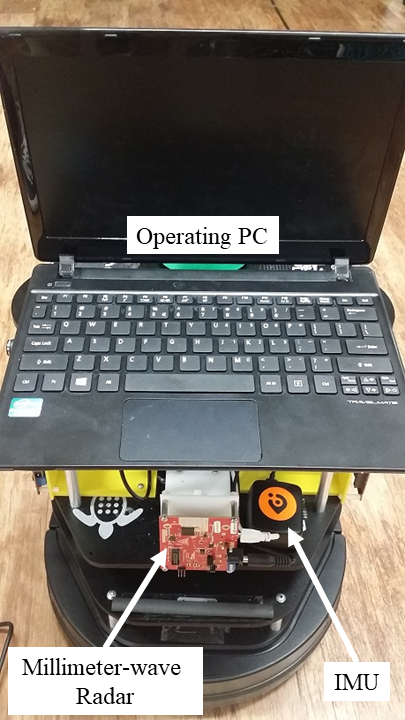}
\caption{}
\label{fig:exp_platform}       % Give a unique label
\end{subfigure}
\begin{subfigure}[t]{0.45\columnwidth}
\includegraphics[width=\textwidth]{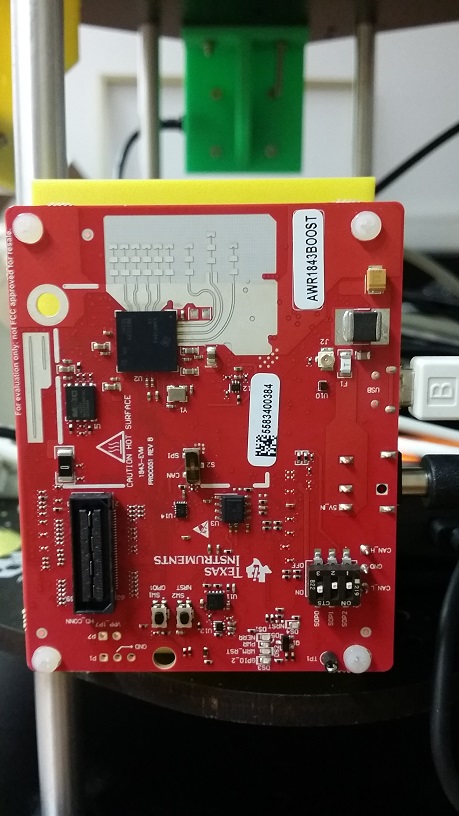}
\caption{}
\label{fig:exp_radar}       % Give a unique label
\end{subfigure}
\caption{\textbf{Experimental setup.} a) Turtlebot2 data collection platform. b) TI AWR1843BOOST model, short-range millimetre-wave radar employed in the experiments.}
\label{fig:exp_setup}
\end{figure}

\begin{figure*}
\centering
\begin{subfigure}[c]{.33\textwidth}
\includegraphics[width=\textwidth]{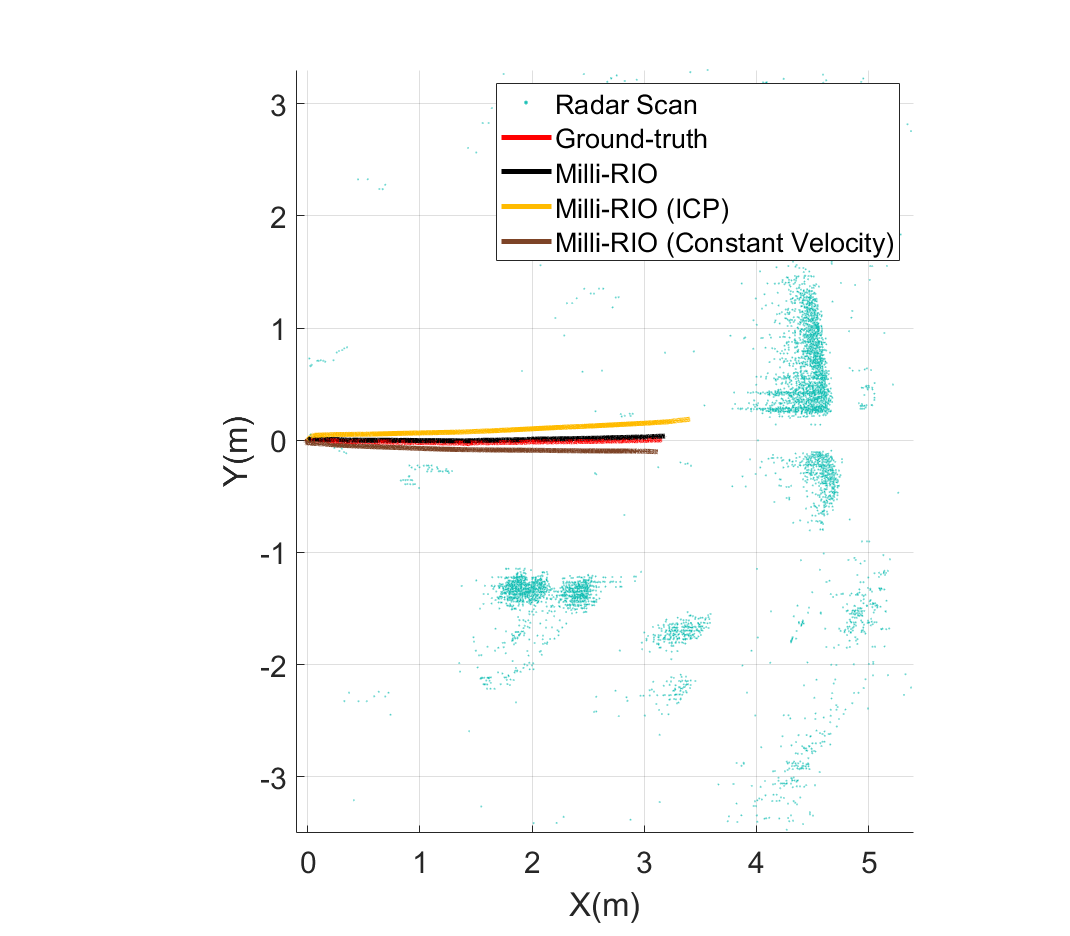}
\caption{Traj.1}
\label{fig:t1_result}       % Give a unique label
\end{subfigure} %
\begin{subfigure}[c]{0.32\textwidth}
\includegraphics[width=\textwidth]{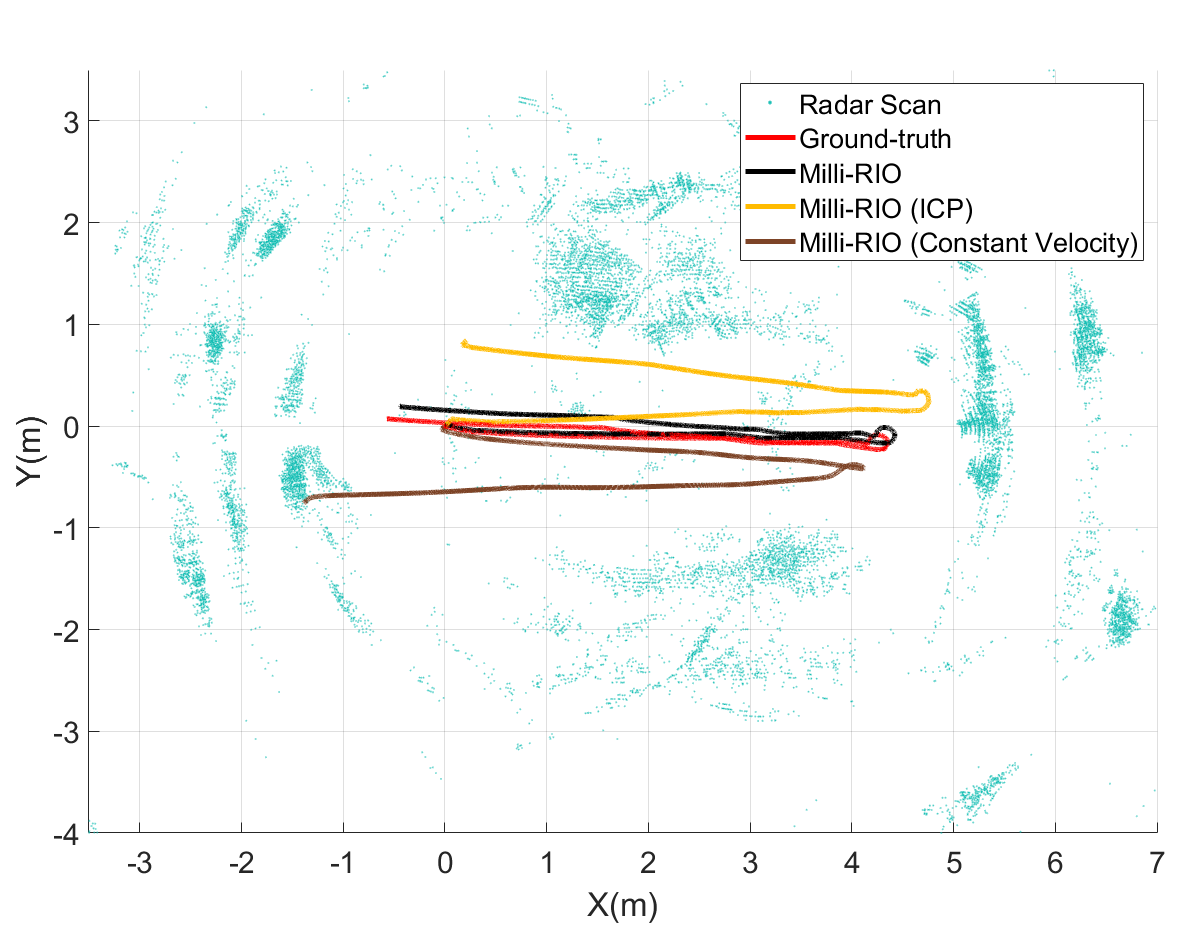}
\caption{Traj.2}
\label{fig:t2_result}       % Give a unique label
\end{subfigure}
\begin{subfigure}[c]{.33\textwidth}
\includegraphics[width=\textwidth]{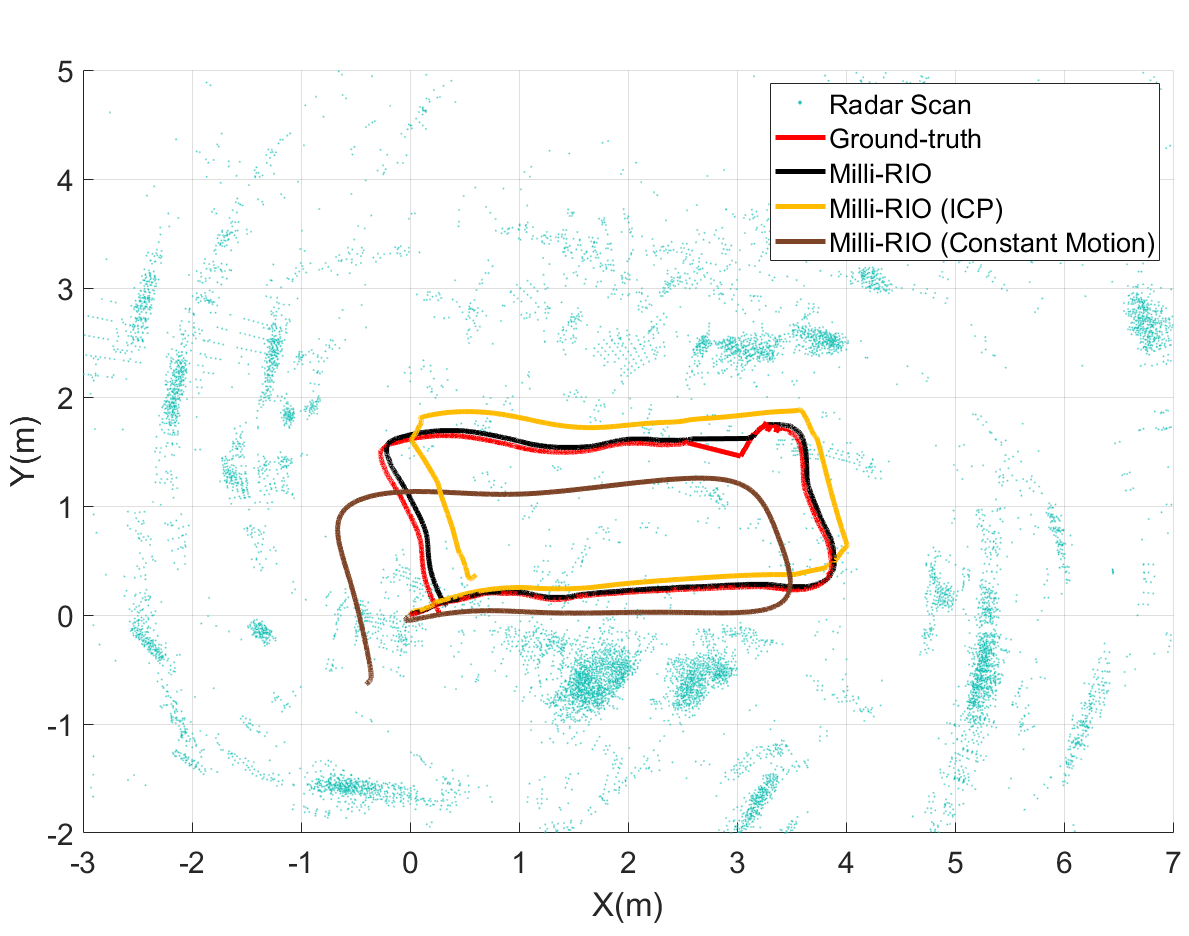}
\caption{Traj.3}
\label{fig:t3_result}       % Give a unique label
\end{subfigure} %
\begin{subfigure}[c]{0.33\textwidth}
\includegraphics[width=\textwidth]{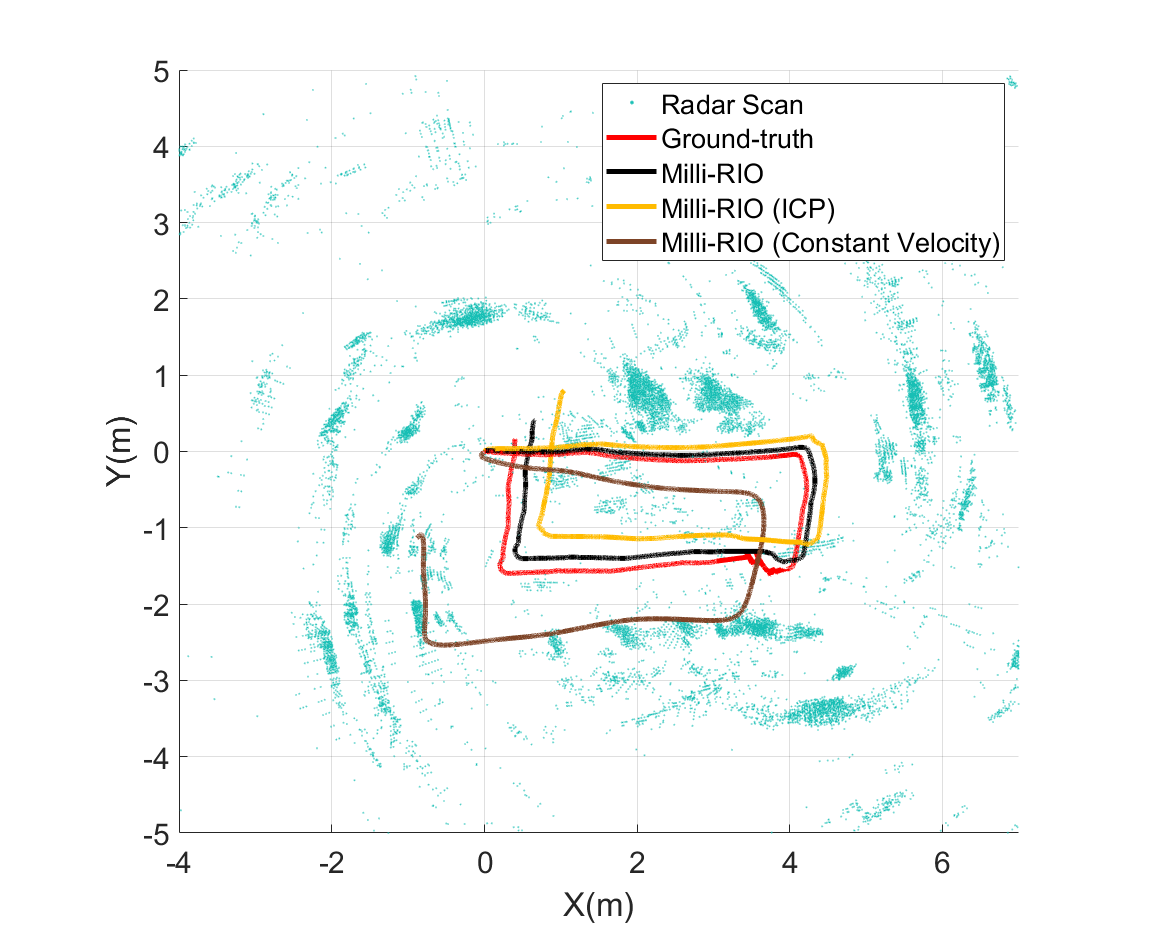}
\caption{Traj.4}
\label{fig:t4_result}       % Give a unique label
\end{subfigure}%
\begin{subfigure}[c]{0.33\textwidth}
\includegraphics[width=\textwidth]{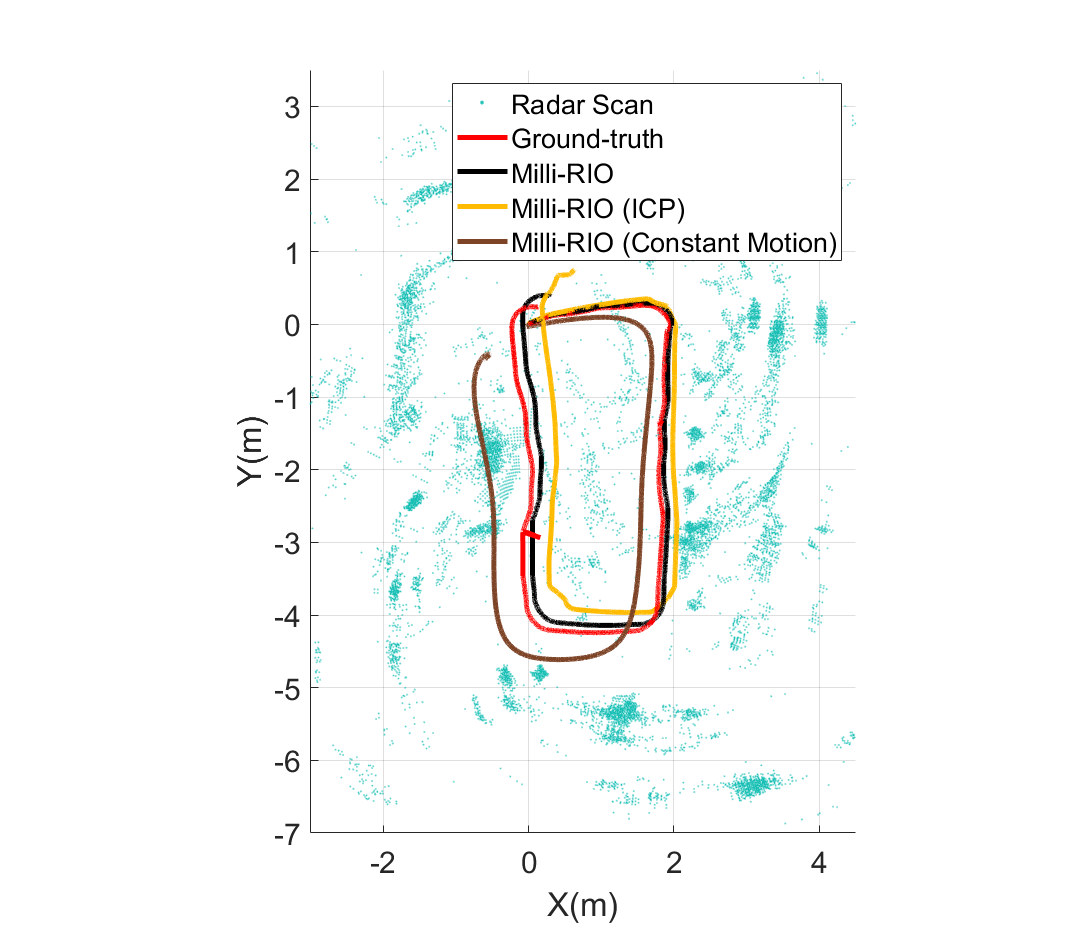}
\caption{Traj.5}
\label{fig:t5_result}       % Give a unique label
\end{subfigure}%
\begin{subfigure}[c]{0.33\textwidth}
\includegraphics[width=\textwidth]{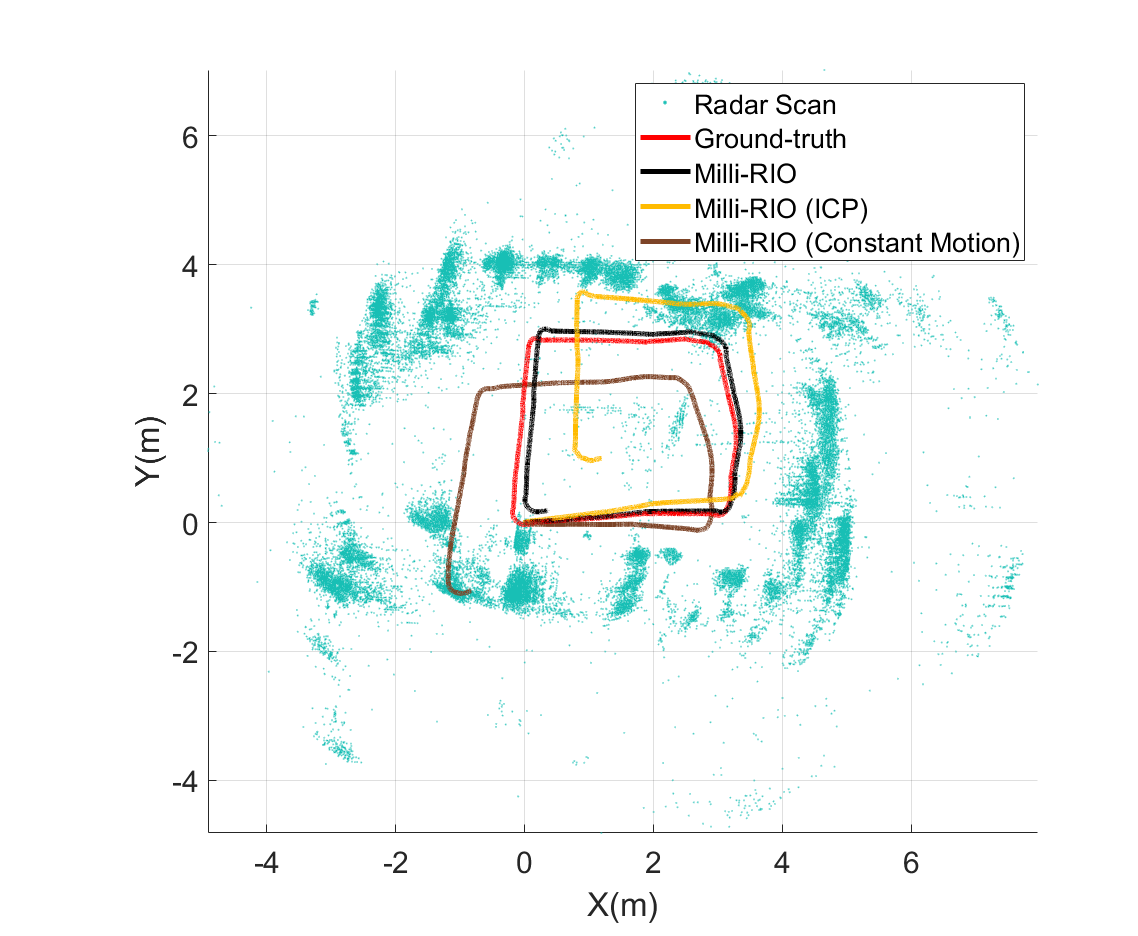}
\caption{Traj.6}
\label{fig:t6_result}       % Give a unique label
\end{subfigure}%
\caption{\textbf{Trajectory estimation results.} Sample trajectories of the moving radar estimated by the proposed method are compared against  ground-truth and traditional approaches. Milli-RIO (ICP) replaces the normal distributions transform (NDT) scan matching module of Milli-RIO with the iterative closest point (ICP) approach. Milli-RIO (Constant Velocity) replaces the recurrent neural network based motion model of Milli-RIO with a constant velocity model. The proposed Milli-RIO architecture outperforms the traditional approaches. The trajectories include various type of motions such as linear and circular motions, sharp turns, and smooth transitions. Point clouds in blue are the registered radar scans.}
\label{fig:t_result}
\end{figure*}

\subsection{RNN-based Motion Model} 
Existing motion estimation methods based on traditional filters have limitations for nonlinear dynamic systems. 
UKF accommodates a wide variety of dynamic models, allowing for highly complex dynamics in the state variables given an accurate motion model. 

In the last decade, deep learning (DL) techniques have exceeded the performance of traditional methods in various domains such as computer vision, speech recognition and natural language processing.
Contrary to these high-level tasks, the motion estimation problem is mainly concerned with the dynamics and the temporal relations across pose sequences coming from different ego-motion algorithms, which can be formulated as a sequential learning problem. 
Unlike traditional feed-forward DL networks, RNNs are very suitable to model the dependencies across time sequences and to create a temporal motion model. 
RNNs represent the current hidden state as a function of arbitrary sequences of inputs by having a memory of hidden states over time and directed cycles among hidden units.
LSTM is a specific implementation of RNN to avoid the vanishing gradient problem, enabling the exploitation of temporal position information for a long time. Thus, LSTM has a higher capacity of learning long-term relations among the pose sequences by introducing memory gates such as input, forget and output gates, and hidden units of several blocks.

The pose estimation of the current time step benefits from information encapsulated in previous time steps. Thus, LSTM is a suitable RNN implementation to formulate the state transition function $f$ in Eq. \ref{eq:predict_x} \cite{turan2018endosensorfusion}.
Our implementation is based on bi-directional LSTM (bi-LSTM) that has a memory function not only for the forward sequences but also for the backward sequences to fully consider the mutual relationship between the sequences.
UKF tracks the 6-DoF pose of moving radar using the transition function modeled by a bi-LSTM network that consists of $256$ LSTM nodes in each direction.
We recorded various trajectories containing rotational and translational motions to train our bi-LSTM network. We use different trajectories for training and testing to ensure the network captures the motion dynamics and avoids overfitting to the training dataset.
We also use batch normalization and dropout layers with a rate of $\alpha=0.25$ to prevent overfitting.
The inputs to the bi-LSTM are accelerometer and gyroscope readings (states) at time step $t-1$, and output labels are 6-DoF poses at time $t$. 
In that way, the bi-LSTM learns the non-linear motion model of the mobile radar.

\section{Experiments and Results}
\label{sec:exps_results}
In this section, we describe our experimental setup and single-chip MMWave radar configurations. We present the spatial and temporal sensor calibration approach employed in our experiment and the details of the dataset creation procedure. Moreover, we show and discuss the evaluation results with quantitative and qualitative metrics. 

\subsection{Low-Cost Millimetre-Wave Scanning Radar}

Single-chip MMWave radar is a promising
solution for low-power, self-monitored, ultra-accurate radar systems.
%The range accuracy and measurement stability of the MMWave radar is high, and the measurement results are stable.
MMWave radar has several advantages such as it provides accurate range measurements, gathers readings at close range, and operates at low peak power. 
Sidelobe radiation sent in unintended directions and multipath reflections that occur when a wave encounters additional reflection points before returning to the receiver antenna cause noise and non-existing object locations in the scan data. Other issues causing noise in the data include phase jitter, saturation, and atmospheric attenuation.

We employ a Texas Instruments AWR1843BOOST model, short-range MMWave scanning radar, which is shown in Fig. \ref{fig:exp_setup}. This radar is attached to a mobile agent, and it continuously transmits and receives frequency modulated radio waves within the maximum angular FoV of $120^\circ$. The power peaks received by the antenna corresponds to a position in the environment, indicating the reflectivity, size and orientation of an object at that position.  
The device is an integrated single-chip MMWave sensor based on FMCW radar technology capable of operation in the 76
to 81 GHz band with up to 4 GHz continuous chirp. The AWR1843 includes a monolithic implementation of a two transmit (TX), four receive (RX)  radio
frequency (RF) components system with built-in PLL and A2D converters. The device also integrates a DSP subsystem, which contains TI C674x DSP for the RF signal processing unit to generate a point cloud of $63$ points per frame. The device includes an ARM R4F-based
processor subsystem, which is responsible for front-end configuration, control and calibration.
The starting frequency $f_c$ of the device is $76$ GHz with $12.5$dBm TX power and $15$dB RX noise figure.
The radar is configured with a bandwidth of $4$ GHz, the ramp slope $S$ of $70$ MHz/$\mu$s, resulting in a range resolution of $4.3$cm, maximum unambiguous range of $22.55$m, and radial velocity of $2.28$m/s with $0.29$m/s resolution.
The number of samples per chirp is $128$ and the frame rate is $20$ frames-per-second (fps).
The radar is placed on the roof of a mobile platform with the axis of the antenna perpendicular to the motion plane (see Fig. \ref{fig:exp_setup}). The platform is typically moved between $0.40$ and $0.60$ m/s; when turning, up to $0.40$ rad/s. The robot is driven through a typical lab environment where it is tracked with a VICON tracking system that provides ground-truth with sub-millimeter accuracy.

\begin{figure}
\centering
\begin{subfigure}{\columnwidth}
\includegraphics[width=\textwidth]{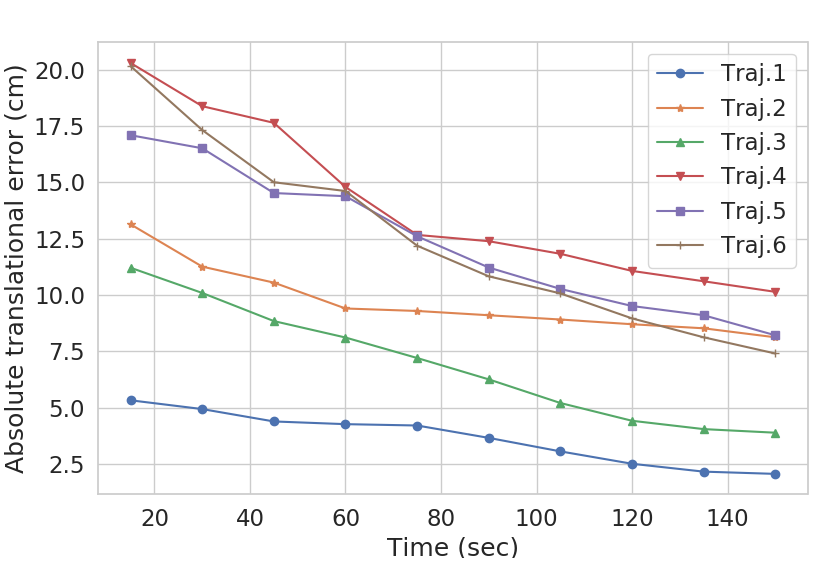}
\caption{Absolute translational error.}
\label{fig:error_trans}       % Give a unique label
\end{subfigure} %
~ 
\begin{subfigure}{\columnwidth}
\includegraphics[width=\textwidth]{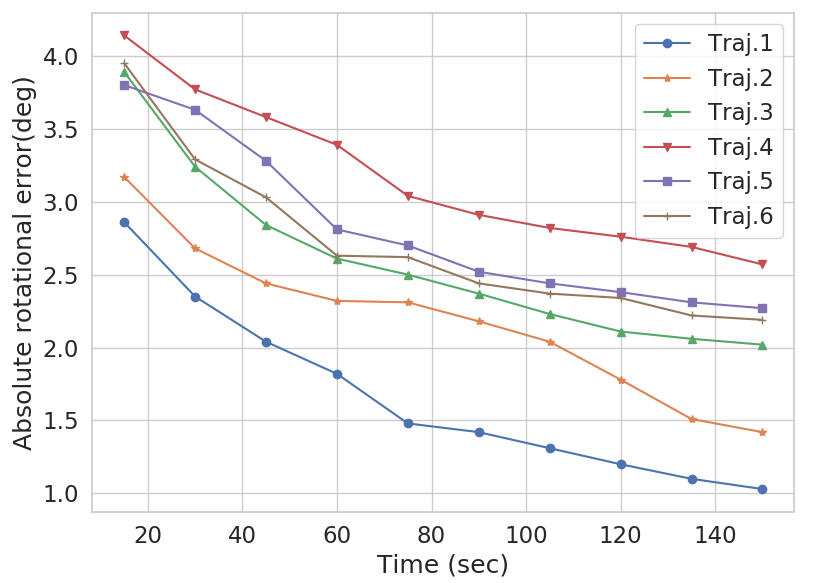}
\caption{Absolute rotational error.}
\label{fig:error_rot}       % Give a unique label
\end{subfigure}
\caption{\textbf{The change of error in time for trajectories in Fig. \ref{fig:t_result}.} Absolute trajectory errors decrease over time because Milli-RIO registers the current point cloud to the accumulated point cloud of the environment, proving the effectiveness of the proposed global alignment approach.}
\label{fig:error_res}
\end{figure}  

\begin{figure}
\centering
\begin{subfigure}{\columnwidth}
\includegraphics[width=\textwidth]{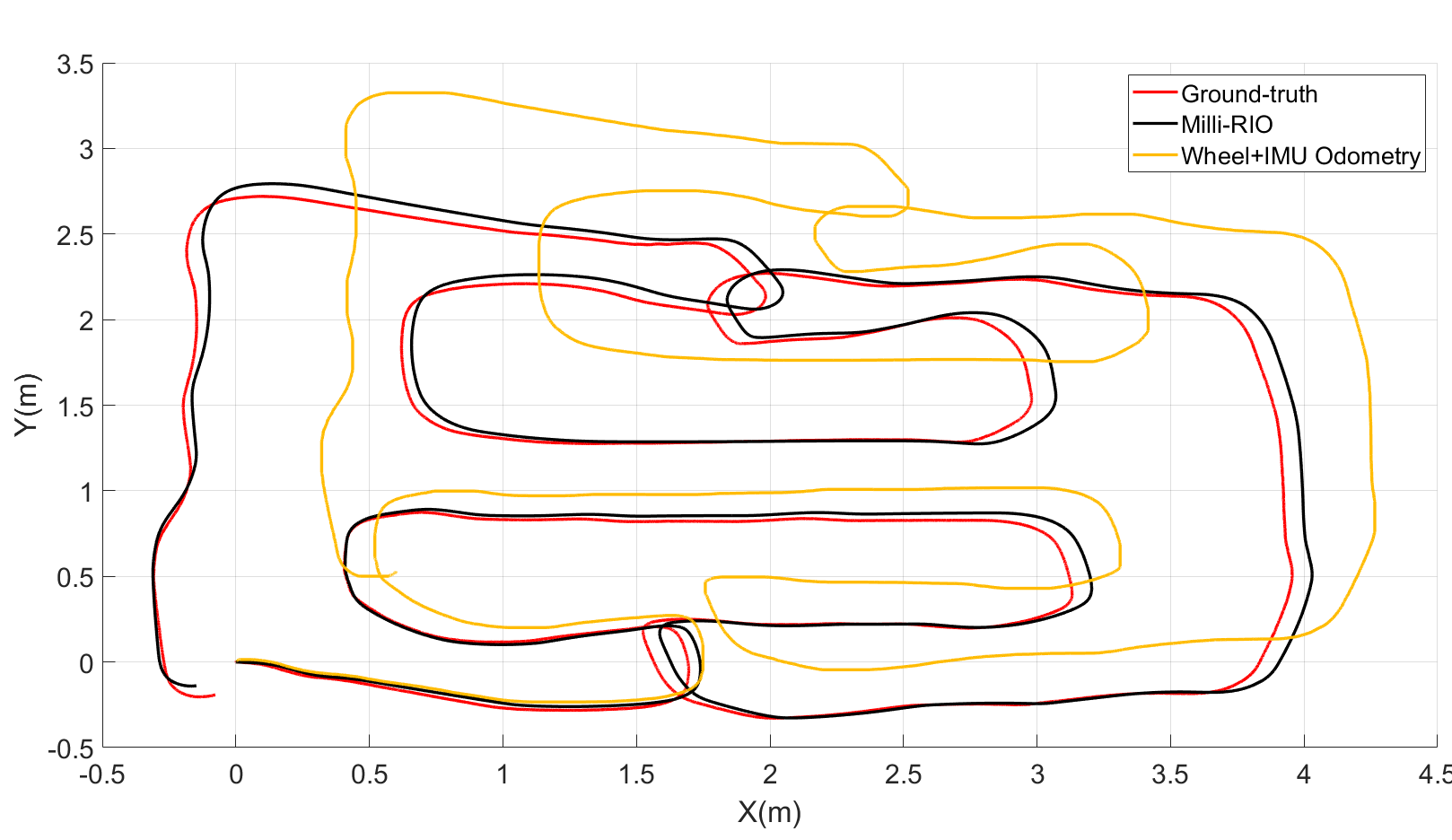}
%\caption{t1}
%\label{fig:error_trans}       % Give a unique label
\end{subfigure} %
~ 
\begin{subfigure}{\columnwidth}
\includegraphics[width=\textwidth]{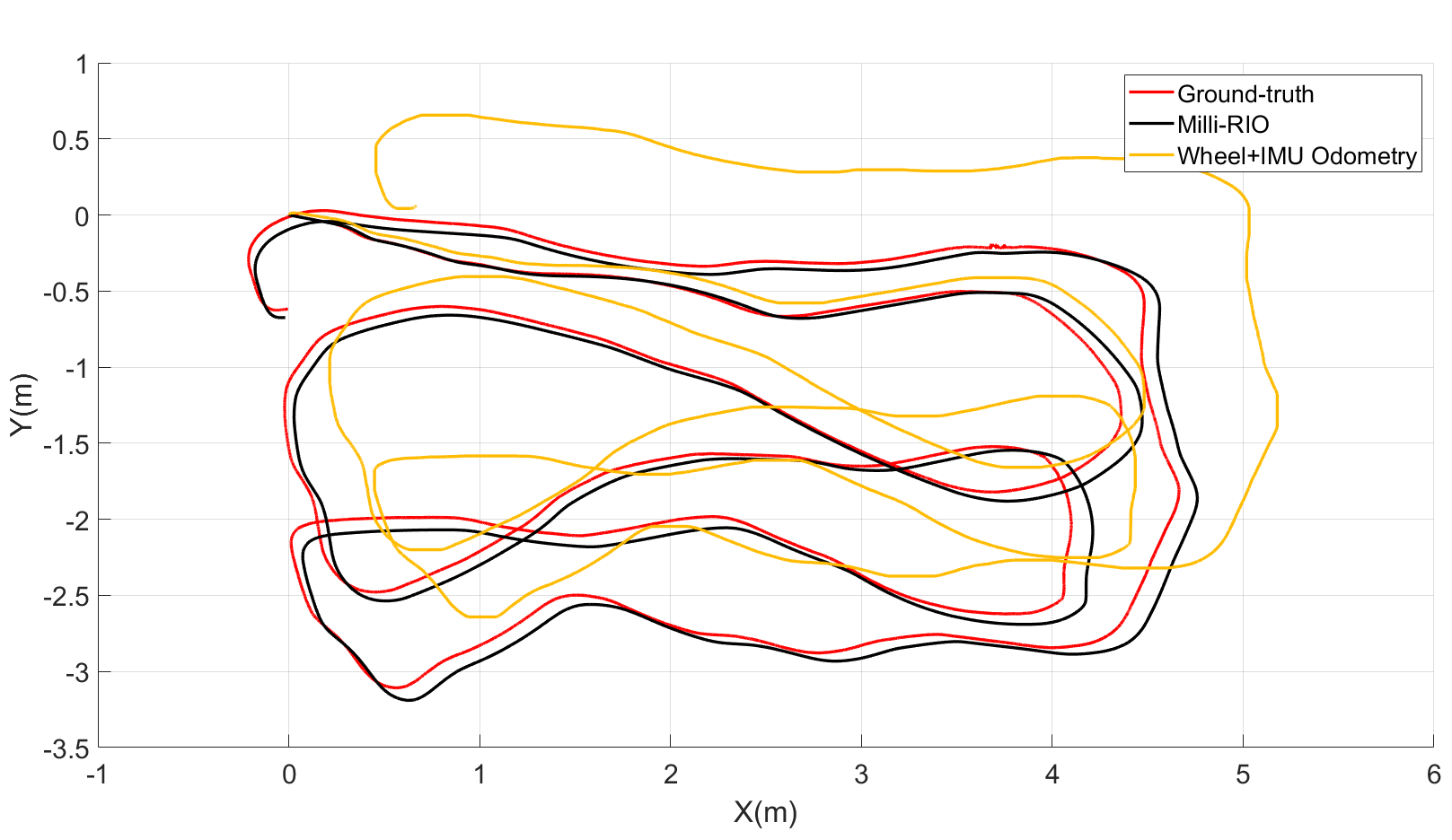}
%\caption{t2}
%\label{fig:traj_long}       % Give a unique label
\end{subfigure}
\caption{\textbf{Comparative odometry estimation performance.} Milli-RIO is resistant to accumulating drift, whereas Turtlebot odometry based on the gyro and motor encoders \cite{turlebotodom} rapidly deviates from the ground-truth.}
\label{fig:traj_long}
\end{figure}  

\subsection{Spatial and Temporal Sensor Calibration}
To calibrate the IMU sensor and MMWave radar with respect to the VICON motion tracking system, we first recorded a sequence with an `$\infty$'-loop. Then, we registered radar scans
using the NDT algorithm. 
To obtain accurate point cloud registration, we placed strongly reflective markers in the environment, which are only used during the calibration session and removed during the experiments. Given pairs of IMU-VICON and radar-VICON trajectories, this problem corresponds to the well-known
hand-eye calibration. We performed hand-eye calibration using the standard approach explained in \cite{horaud1995hand}.

In order to synchronize the sensors, we used the timestamps
of MMWave radar which has a lower fps rate as a reference. 
We collected the information with the closest IMU timestamp to the radar timestamp for a particular frame,
resulting in a worst-case time difference of 5 ms between
IMU and radar data package.
All timestamps were recorded on our host computer using
ROS \cite{quigley2009ros} system clock.

\subsection{Assessment of Odometry Performance}
The dataset is collected in typical office environments, including various types of translational and rotational motions. Such a detailed dataset enables us to evaluate if the proposed method is biased towards certain motion types.
The total path length of the trajectories in Fig. \ref{fig:t_result} is $61.38$m with average trajectory length of $10.23$m, which is recorded in a total time of $913$sec. 
In addition, we record longer trajectories to evaluate the robustness of the proposed method against motion drifts.
The trajectories in Fig. \ref{fig:traj_long} have longer average length of $26.90$m with a total path length of $53.81$m and duration of $628$sec. The trajectories contain both sharp and smooth motions to evaluate the robustness of the proposed approach in indoor environments.

Figure \ref{fig:t_result} illustrates sample trajectories of the moving radar and the corresponding estimated trajectories by the proposed radar-based odometry system.
Figure \ref{fig:t_result} also depicts the overall point cloud registration performed by the proposed approach.
To demonstrate the effectiveness of Milli-RIO architecture, we perform ablation studies, as shown in Fig. \ref{fig:t_result}.
We replace the normal distributions transform (NDT) scan matching module of Milli-RIO with the standard ICP approach, called Milli-RIO (ICP). Although Milli-RIO (ICP) estimates the full trajectories without tracking failure, it is prone to motion drift and deviates from the ground-truth.
Furthermore, we replace the RNN based motion model of Milli-RIO with the standard constant velocity model in UKF, called Milli-RIO (Constant Velocity). However, Milli-RIO (Constant Velocity) suffers from a significant drift over time. Although it has a slightly better performance on trajectories following a line, it performs poorly on trajectories that contain sharp turns and rotations.
The proposed Milli-RIO architecture outperforms traditional approaches.
In addition, we removed the radar odometry from the motion estimation pipeline to test the contribution of the IMU sensor to the overall motion estimation performance. However, the filter rapidly diverges and the motion estimation fails after few iterations due to the accumulated quadratic error in the double integration of unconstrained IMU bias and noise, proving the effectiveness of radar odometry.

In Fig. \ref{fig:error_res}, we display both the translational and rotational ATE (absolute trajectory error) in $cm$ and $deg$, respectively. The translational and rotational error decreases over time because Milli-RIO registers the current point cloud to the accumulated point cloud of the environment. Such a global alignment approach is more effective than local alignment due to better data association.
Table \ref{tab:ate_trans} and \ref{tab:ate_rot} quantitatively shows ATE results in terms of mean, median, standard deviation and root mean square error (RMSE).

%As shown in Fig.\ref{fig:t_result} and Tables \ref{tab:ate_trans} and \ref{tab:ate_rot}, high concentration in the clustered regions of point clouds results in lower trajectory error. Similarly, scattered point clouds reduce the performance of point cloud registration and, thus, cause higher trajectory errors.    
%One can see the successful pose estimation of the user-defined trajectories with minimal deviations on the order of centimeter-scale in both Fig. \ref{fig:t_result} and Table \ref{tab:ate_trans}. 
%The signal attenuation resulting from radiation through a different medium creates unstable points in radar scans, which causes deviations from the ground-truth trajectories. 

We further compare Milli-RIO with an off-the-shelf odometry method, fusing wheel odometry with the inertial navigation system.
Figure \ref{fig:traj_long} shows the odometry results of Milli-RIO on longer trajectories, comparing with the ground-truth and Turtlebot odometry based on the gyro and motor encoders \cite{turlebotodom}.
The evaluations prove that the proposed method is resistant to accumulating drift even on long indoor trajectories that contain complex motions.
A video demonstration of Milli-RIO is available online\footnote{https://youtu.be/VoPKrC8st8I}.

\begin{table}
\small
\begin{center}
\setlength{\tabcolsep}{5.0pt}
\begin{tabular*}{1.0\linewidth}{c|c|c|c|c|c|c}
\hline
Error (cm) & Traj.1 & Traj.2 & Traj.3 & Traj.4 & Traj.5 & Traj.6 \\
\hline
Mean & $2.57$ & $9.06$ & $4.81$ & $12.39$ & $10.96$ & $10.28$\\
Median & $2.54$ & $9.09$ & $4.67$ & $12.27$ & $9.06$ & $10.29$\\
Std. & $1.49$ & $5.28$ & $2.93$ & $7.59$ & $5.51$ & $8.26$\\
RMSE & $2.97$ & $10.48$ & $5.63$ & $10.22$ & $10.98$ & $13.50$\\
\hline
\end{tabular*}
\end{center}
\caption{Translational ATE (absolute trajectory error) results for MILLI-RIO.}
\label{tab:ate_trans}
\vspace{-2ex}
\end{table}

\begin{table}
\small
\begin{center}
\setlength{\tabcolsep}{5.0pt}
\begin{tabular*}{1.0\linewidth}{c|c|c|c|c|c|c}
\hline
Error (deg) & Traj.1 & Traj.2 & Traj.3 & Traj.4 & Traj.5 & Traj.6 \\
\hline
Mean & $1.38$ & $1.93$ & $2.43$ & $2.87$ & $2.60$ & $2.37$\\
Median & $1.25$ & $1.76$ & $2.27$ & $2.72$ & $2.54$ & $2.69$\\
Std. & $1.01$ & $1.33$ & $1.86$ & $2.17$ & $1.79$ & $2.15$\\
RMSE & $1.49$ & $1.60$ & $2.09$ & $2.35$ & $2.20$ & $2.55$\\
\hline
\end{tabular*}
\end{center}
\caption{Rotational ATE (absolute trajectory error) results for MILLI-RIO.}
\label{tab:ate_rot}
\vspace{-2ex}
\end{table}

\section{Conclusion}
\label{sec:conclusion}

In this paper, we introduced an accurate and robust radar-IMU motion
estimation system that achieves centimeter accuracy and demonstrates the
effectiveness of MMWave radars for indoor localization. As an onboard low-cost radar sensor, 
the successful implementation of MMWave radar odometry improves the reliability and
versatility of mobile systems. Our method stands out because
it is not only dependable and accurate but also straightforward and intuitive without a need for a hand-engineered motion model. In the future, we plan to incorporate a robust 3D map reconstruction module into the pipeline.

\bibliographystyle{IEEEtran}
\bibliography{IEEEabrv,mybibfile}

\begin{IEEEbiographynophoto}{Yasin Almalioglu} is currently enrolled as a DPhil (PhD) student in computer science at the University of Oxford since January 2018.
He received the BSc degree with honors in computer engineering from Bogazici University, Istanbul, Turkey, in 2015. He was a research intern at CERN Geneva, Switzerland and Astroparticle and Neutrino Physics Group at ETH Zurich, Switzerland, in 2013 and 2014, respectively. He was awarded the Engin Arik Fellowship in 2013. He received the MSc degree with high honors in computer engineering from Bogazici University, Istanbul, Turkey, in 2017. 
\end{IEEEbiographynophoto}

\begin{IEEEbiographynophoto}{Mehmet Turan} received his diploma degree from RWTH Aachen University, Germany in 2012 and his PhD degree from ETH Zurich, Switzerland in 2018.
Between 2013-2014, he was a research scientist at UCLA (University of
California Los Angeles). Between, 2014-2018 he was a research scientist
at Max Planck Institute for Intelligent Systems and between 2018-2019,
he was a postdoctoral researcher at the Max Planck Institute for
Intelligent Systems. He received the DAAD (German Academic Exchange
Service) fellowship between years 2005–2011 and Max Planck fellowship
between 2014–2019. He has also received Max Planck-ETH Center for
Learning Systems fellowship between 2016–2019. Currently, he is a faculty at the Institute of Biomedical Engineering, Bogazici University, Turkey.  
\end{IEEEbiographynophoto}

\begin{IEEEbiographynophoto}{Chris Xiaoxuan Lu} is a post-doctoral fellow at University of Oxford, working on EPSRC Programme Grant Project “Mobile Robotics: Enabling a Pervasive Technology of the Future”.
He obtained his Ph.D. degree in Computer Science at Oxford University. His Ph.D. study was generously funded by Google DeepMind. Prior to joining Oxford, he received his MEng degree in 2015 from Nanyang Technological University (NTU).
\end{IEEEbiographynophoto}

\begin{IEEEbiographynophoto}{Niki Trigoni} is a Professor at the Oxford University Department of Computer Science and a fellow of Kellogg College. She obtained her DPhil at the University of Cambridge (2001), became a postdoctoral researcher at Cornell University (2002-2004), and a Lecturer at Birkbeck College (2004-2007). At Oxford, she is currently Director of the EPSRC Centre for Doctoral Training on Autonomous Intelligent Machines and Systems, a program that combines machine learning, robotics, sensor systems and verification/control. She also leads the Cyber-Physical Systems Group, which is focusing on intelligent and autonomous sensor systems with applications in positioning, healthcare, environmental monitoring and smart cities.
\end{IEEEbiographynophoto}

\begin{IEEEbiographynophoto}{Andrew Markham} is an Associate Professor and he works on sensing systems, with applications from wildlife tracking to indoor robotics to checking that bridges are safe. He works in the cyber-physical systems group. He designs novel sensors, investigates new algorithms (increasingly deep and reinforcement learning-based), and applies these innovations to solving new problems. Previously he was an EPSRC Postdoctoral Research Fellow, working on the UnderTracker project. He obtained his Ph.D. from the University of Cape Town, South Africa, in 2008, researching the design and implementation of a wildlife tracking system using heterogeneous wireless sensor networks.
\end{IEEEbiographynophoto}

\end{document}